\documentclass[twocolumn,twoside,slac_two]{revtex4}
\usepackage{graphicx}
\usepackage{fancyhdr}
\newcommand{\sgr}{\mbox{SGR\,J1550--5418}}
\newcommand{\psr}{\mbox{PSR\,J1550--5418}}
\newcommand{\esrc}{\mbox{1E\,1547.0--5408}}
\newcommand{\trig}{$T_0$}
\newcommand{\degrees}{\ensuremath{^\circ}}

\begin{document}

\title{Fermi/Gamma-ray Burst Monitor detection of  \sgr}

\author{Yuki~Kaneko}
\author{Ersin~G\"o\u{g}\"u\c{s}}
\affiliation{Sabanc\i~University, Orhanl\i-Tuzla, \.Istanbul 34956, Turkey}
\author{Chryssa~Kouveliotou}
\affiliation{Space Science Office, VP62, NASA/Marshall Space Flight Center, Huntsville, AL 35812, USA}
\author{Jonathan~Granot}
\affiliation{Centre for Astrophysics Research, University of Hertfordshire, College Lane, Hatfield AL10 9AB, UK}
\author{Enrico~Ramirez-Ruiz}
\affiliation{Department of Astronomy and Astrophysics, University of California, Santa Cruz, CA 95064, USA}
\author{On behalf of GBM Magnetar Team}

\begin{abstract}
\sgr~exhibited three active bursting episodes in 2008 October and in 
2009 January and March, emitting hundreds of typical Soft Gamma Repeater (SGR) 
bursts in soft gamma-rays.  The second episode was especially intense, 
and our untriggered burst search on {\it Fermi}/GBM data (8$-$1000~keV) 
revealed $\sim$450 bursts emitted over 24 hours during the peak of this 
activity. Using the GBM data, we identified a $\sim$150-s-long enhanced 
persistent emission during 2009 January 22 that exhibited intriguing 
timing and spectral properties: (i) clear pulsations up to $\sim$110~keV 
at the spin period of the neutron star, (ii) an additional (to a power-law) 
blackbody component required for the enhanced emission spectra with $kT 
\sim 17$~keV, (iii) pulsed fraction that is strongly energy dependent and 
highest in the 50$-$74~keV energy band.  A total isotropic-equivalent 
energy emitted during this enhanced emission is estimated to be $2.9 
\times 10^{40}(D/5\,{\rm kpc})^2$~erg.  The estimated area of the 
blackbody emitting region of $\approx 0.046(D/5\,{\rm kpc})^2\;{\rm km}^2$ is the smallest 
``hot spot" ever measured for a magnetar and most likely corresponds to 
the size of magnetically-confined plasma near the neutron star surface.

\end{abstract}

\maketitle

\thispagestyle{fancy}

\section{Introduction}

\esrc~was observed with the X-ray Multi-Mirror Mission (XMM-Newton) in 
2004 as a magnetar candidate, selected for its galactic plane location
and its relatively soft magnetar-like spectrum as seen with the
Advanced Satellite for Cosmology and Astrophysics (ASCA) during their
Galactic plane survey \cite{sug01}. Although no period was detected
in the original and follow-up XMM observations, its
positional coincidence with an extended galactic radio source
G327.24-0.13 (possibly a supernova remnant) also suggested that \esrc~is a magnetar \cite{gel07}.
The subsequent discovery
in radio observations of a spin period of 2.07~s and a period
derivative of 2.3 $\times$~10$^{-11}$~s~s$^{-1}$ led to an estimated
dipole surface field of $B\sim2.2\times10^{14}$ G and confirmed the
source's magnetar nature; the source was also renamed as
\psr~\cite{cam07}. Its period makes \esrc~the fastest rotating
magnetar; the source is also one of the only two that emit in radio
wavelengths \cite{hal05, cam06}. The distance of the source has been
estimated to be $\sim$4$-$9~kpc using various methods \cite{cam07, gel07}, including the most recent estimate of 4$-$5~kpc \cite{tie09}.
Throughout this paper we use $D_5 = D/5$~kpc as the source
distance measure.

On 2008 October 3, \esrc~entered an episode of X-ray activity,
emitting several typical SGR-like bursts over the next 7 days. During
this period, 22 short duration bursts were observed with the Gamma-ray
Burst Monitor (GBM) on board the {\it Fermi} Gamma-ray Space
Telescope \cite{von10}.

The source entered a second period of extremely
high X-ray burst activity on 2009 January 22 \cite{mer09}.  During the first 24~hours of this bursting episode, 
the {\it Fermi}/GBM triggered on the source 41 times:
the number of triggers was limited only by the instrument's capability
and did not reflect the actual number of bursts emitted by the
source. In fact, our on-ground search for untriggered events revealed
a total of $\sim$450 bursts during this 24 hour period: an unusually
high burst frequency from a single source \cite{van10}. Based on this SGR-like behaviour,
we renamed the source as \sgr~\cite{kou09}.

We discovered an enhancement of the persistent emission lasting $\sim$150~s by examining 
 the data from the first GBM trigger on January 22.
Closer inspection
of this enhancement in various energy ranges revealed periodic
oscillations with a period consistent with the spin period of \sgr. Here we
present a detailed temporal and spectral analysis of this
enhanced emission. In \S\ref{sec:obs}, we briefly describe our
observations and the GBM instrument and data types. We present our
temporal analysis results in \S\ref{sec:tempo}, and our spectral
studies in \S\ref{sec:spec}. Finally we discuss the physical
implications of our discovery in \S\ref{sec:sum}.
Complete descriptions of our analysis and discussion are presented in Kaneko et al. (2010) \cite{kan10}.

\section{Instrumentation and Data}\label{sec:obs}

In trigger mode, GBM provides three types of data for each of 12 NaI and 2 BGO detectors; CTIME
Burst, CSPEC Burst, and Time Tagged Event (TTE) data
\cite{mee09}. The CTIME Burst data have a time resolution of 64~ms
with rather coarse spectral information (8~energy channels). The CSPEC
Burst data provide high-resolution spectra (128~energy channels)
collected every 1.024~s. Both CTIME Burst and CSPEC Burst accumulate
data for $\sim$600~s after a trigger. The TTE data provide time-tagged
photon event lists for an accumulation time of 330~s, starting 30 s
prior to the trigger time; this data type provides a superior temporal
resolution down to 2$\mu$s at the same spectral resolution as the
CSPEC Burst data.

\begin{figure}[tb]
\includegraphics[width=0.5\textwidth]{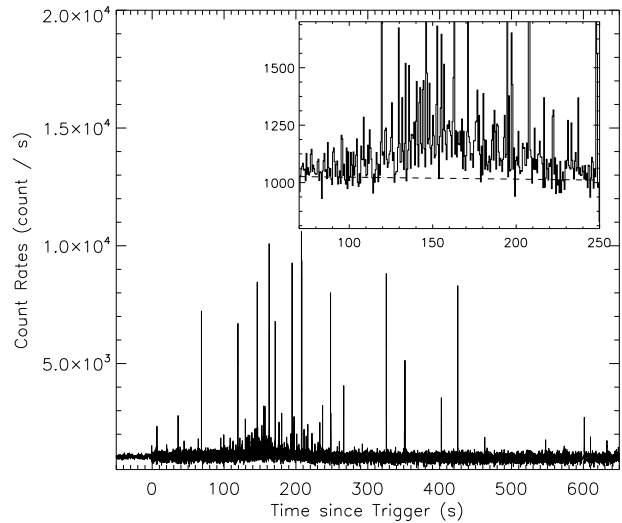}
\caption{GBM lightcurve of \sgr~in 12$-$293~keV.  An enlarged view of the pulsed, enhanced emission is
shown in the inset.  The dashed line indicates the background level.}
\label{fig:lc_ch14}
\end{figure}

The first GBM trigger at the onset of the second active episode from
\sgr~was on 2009 January 22 at 00:53:52.17 UT (= \trig, GBM trigger
number 090122037). In the 600~s of the trigger readout we detected
many individual short bursts using our on-ground untriggered burst
search algorithm.  To accept an event as an untriggered burst, we
required excess count rates of at least 5.5$\sigma$ and 4.5$\sigma$ in
the first and second brightest detectors, respectively, in the 10$-$300~keV energy range. We used CTIME data in both continuous (256~ms time
resolution) and Burst mode (64~ms resolution).  Subsequently, we inspected energy-resolved burst morphology and compared each
detector zenith angle to the source for all 12 detectors, to determine
whether the events originated from \sgr. In total we identified about
a dozen very bright bursts and over 40 less intense bursts within
600~s after \trig~(see Figure \ref{fig:lc_ch14}). During the same
trigger readout we also discovered an enhancement in the underlying
persistent emission starting at approximately \trig$+ 70$~s and
lasting for $\sim$150~s (see inset of Figure \ref{fig:lc_ch14}).

In the analysis presented here, we have exclusively used data from
NaI\,0, which initially had the smallest detector zenith angle to the source (15\degrees) and to which the
source was visible through most of the enhanced emission.
We also note that we checked the LAT data (20~MeV$-$300~GeV) of the entire
day for associated high-energy gamma-ray emission, but found no
evidence of high-energy photons originating from the direction of
\sgr.

\section{Temporal Properties} \label{sec:tempo}

\subsection{Timing Analysis}

\begin{figure}
\includegraphics[width=0.49\textwidth]{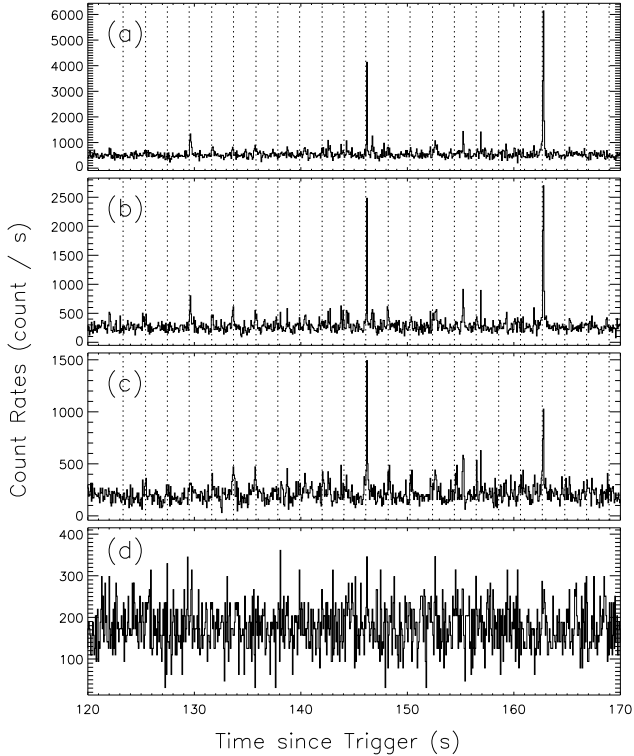}
\vspace{0pt}
\caption{Lightcurve of \sgr~in various energy ranges;
(a) 12$-$27~keV, (b) 27$-$50~keV, (c) 50$-$102~keV, and (d)
102$-$293~keV.  The pulsations are most prominent between 50$-$102~keV (panel
c, starting at $\sim$130~s). The bursts have not been removed here from the data. 
The dashed lines in panels (a) to (c) indicate the times of the pulse maxima.}
\label{fig:lc_pulse}
\end{figure}

During our search for untriggered events in the first trigger interval
of 2009 January 22 from \sgr, we found strong apparent periodic
modulations in the enhanced emission period from \trig+130 to 160 s in
the 50$-$102 keV data of detector NaI\,0 (see panel (c) of
Figure~\ref{fig:lc_pulse}). 
This is the first time to our knowledge
that pulsations unrelated to a giant flare from a magnetar were
clearly seen in the persistent emission of an SGR, in energies up to 100~keV.
To search for a coherent pulse period, we performed a timing analysis
over the entire enhancement interval.  We first eliminated the times
of all the bursts found via our untriggered burst search and converted
the remaining burst-free intervals to the solar system barycenter. 
We then generated a Lomb-Scargle periodogram
\cite{lom75,sca82} over a range of periods from 0.1~s to 10~s using
CTIME Burst data in the 50$-$102~keV band. We found a very significant
signal with a Lomb power of 72.6 (chance occurrence probability, P$_c
\simeq10^{-16}$) at a period of 2.0699~$\pm$~0.0024~s, which is
consistent with the spin period of \sgr. Further, to confirm our
detection, we also employed the Z$^2$$_m$ test (with m =
2) \cite{buc83} on the burst-free and barycentered TTE data. We find a
coherent signal (with a Z$^2$$_{m=2}$ power of 266, P$_c
\simeq10^{-23}$) at the same period. Our spin period measurement is
consistent with the one found for \sgr~using contemporaneous X-ray
data ({\it Swift}/XRT) \cite{kui09, isr09} and radio data \cite{bur09}.  Therefore, we clearly confirm
with the detection of these hard X-ray pulsations that the enhanced
persistent emission seen in the inset of Figure~\ref{fig:lc_ch14}
originates from \sgr.

Next we searched in the enhanced persistent emission for evolution in
the intensity of the pulsations using a sliding boxcar technique. We
found that the pulsed signal peaks over a 90~s interval, from
\trig$+120$ to $210$~s, which encompasses the peak of the enhancement.
Finally, we searched for any other intervals exhibiting pulsed
emission in the burst-free continuous CTIME data of 2009 January 22
and during the four subsequent days, using a sliding boxcar of 120 s
with 10 s steps. We did not find any additional
statistically-significant pulsed emission.
For the entire search and for all the timing analysis reported here,
we used more precise spin ephemeris obtained by contemporaneous
{\it Swift}/XRT, Chandra, XMM-Newton and Suzaku observations \cite{isr10}.

\subsection{Pulse Profiles}\label{sec:pprofile}

\begin{figure}
\includegraphics[width=0.49\textwidth]{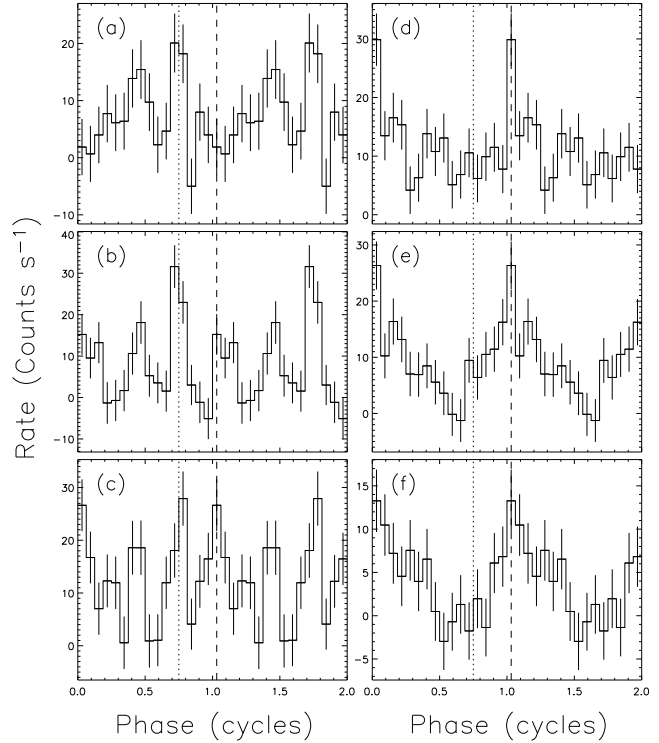}
\vspace{0pt}
\caption{Pulse profiles of \sgr~in equal logarithmic energy intervals;
(a) 10$-$14~keV, (b) 14$-$22~keV, (c) 22$-$33~keV, (d) 33$-$50~keV,
(e) 50$-$74~keV, and (f) 74$-$110~keV.  Two cycles are plotted
for clarity. The vertical dotted and dashed lines are explained in \S\ref{sec:pprofile}.
\label{fig:pptte}}
\end{figure}

To investigate the evolution of the pulse profiles with energy, we
folded the burst-free TTE data spanning 120~s (from \trig+100~s to
\trig+220~s, which includes the strongest pulsation period as found above) with the 
spin ephemeris of \sgr.  We estimated the background level using the data 
segment between \trig~to \trig+60~s. Figure~\ref{fig:pptte} shows the 
source pulse profiles during the enhanced emission interval in six energy bands 
that have the same logarithmic width. The pulse profiles above 110~keV are
consistent with random fluctuations, and thus not shown.

Figure \ref{fig:pptte} indicates that the \sgr~pulse profiles in the
three lowest energy bands are most likely complex (multi-peaked).
While the two lowest energy band profiles are dominated by the
structure around phase 0.7-0.8 (indicated by the dotted lines in
Figure \ref{fig:pptte}), in the 14$-$22~keV band we see the emergence
of another structure around phase 0.0 (indicated by the dashed lines
in Figure \ref{fig:pptte}). This pulse becomes equally prominent in
the 22$-$33 keV range and then dominates in the 33$-$50 keV band. The
pulse profile changes remarkably in the 50-74 keV band, which is the
most statistically significant of all the energy bands investigated,
and is distinguished by a broad structure that peaks at around phase
0.0. The 74$-$110~keV profile resembles the 50-74 keV one.  As noted
above, the pulse profile above 110~keV is consistent with random
fluctuations.  Therefore, our results set an observed upper
energy bound of 110~keV for the hard X-ray pulsations in \sgr~during
this enhanced emission episode.

\subsection{Pulsed Fraction}

\begin{figure}
\includegraphics[width=0.5\textwidth]{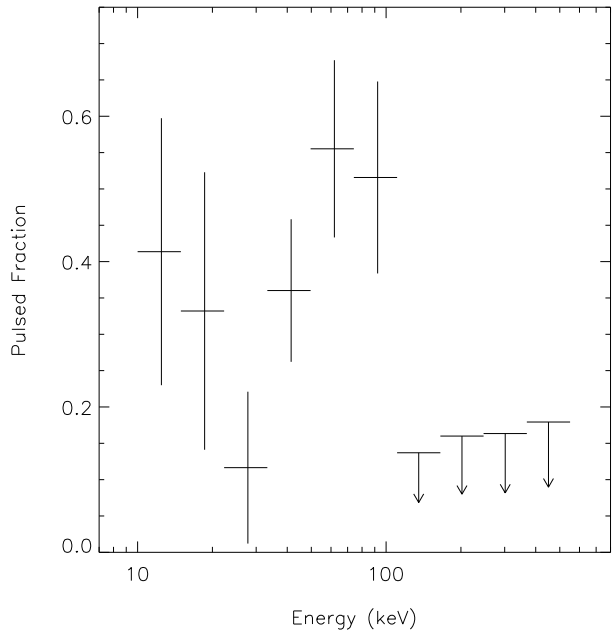}
\caption{Evolution of RMS pulsed fraction of \sgr~as a function of energy.  Uncertainties are 1$\sigma$.
The energy bands are the same as those used in Figure~\ref{fig:pptte}.
\label{fig:pfall}}
\end{figure}

We computed the RMS pulsed fraction using a Fourier based approach as
described in \cite{woo07}. In summary, we take the Fourier transform
of each pulse profile, then we calculate the RMS pulsed flux by taking
the Fourier coefficients of up to third harmonic into account, and
finally obtain the pulsed fraction values by dividing the RMS pulsed
flux by the phase-averaged flux. In Figure~\ref{fig:pfall}, we show
the pulsed fraction spectrum of \sgr~in the same energy bands as in
Figure~\ref{fig:pptte}.
Although marginally significant, there is an indication of a minimum
in the RMS pulsed fraction around $\sim$30~keV. The RMS reaches its
maximum value of 0.55 $\pm$ 0.12 in the 50$-$74 keV band, and then
dips below detection at energies greater than $\sim$110~keV.

\section{Spectral Properties}\label{sec:spec}
\subsection{Time-Integrated and Time-Resolved Spectral Analysis}

We analyzed time-integrated and time-resolved spectra of the enhanced
emission, using the CSPEC Burst data (8.6$-$897~keV).
Similar to the timing analysis, we excluded all bursts identified with the
untriggered search within the enhancement period. 

Since the Detector Response Matrices (DRMs) of GBM are time dependent
due to the continuous slewing of the spacecraft, a DRM should be
generated for every 2-3 degrees of slewing (corresponding to every
$\sim$20$-$50~s of data). 
Therefore, we generated DRMs for every 50~s starting from \trig, for this analysis. 

\begin{figure}
\includegraphics[width=0.49\textwidth]{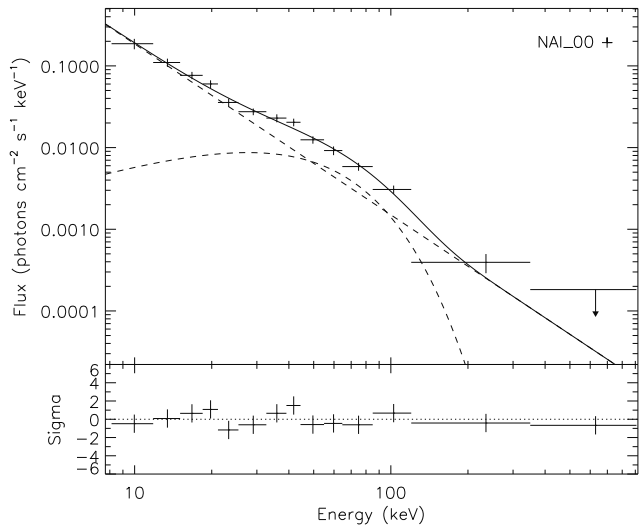}
\caption{The photon spectrum of the time interval \trig+122 to 169~s.  The blackbody and power-law components are shown separately with dashed curves.
The data are binned for display purpose only. A 3$\sigma$ upper limit is shown for the highest energy bin.}
\label{fig:nufnu}
\end{figure}

We found clear evidence for spectral curvature below 100~keV in the
time-integrated spectrum of the entire burst-free enhancement period
(72$-$248~s); a single power law thus resulted in a very poor fit. We
employed five other spectral models; cut-off power law, power law +
blackbody, optically-thin thermal bremsstrahlung, and single/double
blackbody. We found that the time-integrated spectrum is best
described by a power law + blackbody (see Figure \ref{fig:nufnu}). All
other spectral models did not provide better fits mainly because they
failed to fit the lower energy excess $\lesssim10$~keV. The best-fit
power-law index and temperature of the additional blackbody are
shown in Table~\ref{tab:spec_result}. Adding a blackbody (with $kT
=18~\pm$ 4~keV) to a power law resulted in the most significant
improvement in Cash statistics \cite{cas79} over a single power law
($\Delta$C-stat = 13.5 for 2 degrees of freedom, corresponding to an
improvement of 3.25$\sigma$).

\begin{table}[b]
\caption{Spectral parameters of the enhanced persistent emission period of \sgr.
1-$\sigma$ uncertainties are shown in parentheses.}  
\tabcolsep=2pt
\scriptsize
\begin{center}
\vspace{-2ex}
\begin{tabular}{ccccc}
\hline \hline \\[-2ex]
Time & Power\,Law &Blackbody &   \multicolumn{2}{c}{Energy Flux$^{^*}$}  \\
\cline{4-5}
since \trig & Index & $kT$  & Power\,Law & Blackbody \\
(s) & & (keV)  & \multicolumn{2}{c}{{(10$^{-8}$ ergs/cm$^{2}$-s)}}  \\
\hline 
 72$-$248		& $-$2.06 (0.10)	&17.7 (3.8)	& 5.30 (2.37)	& 1.22 (0.28)	\\
\hline									
74$-$117		& $-$2.15 (0.17)	& No BB		& 2.85 (3.30)	& $-$		\\
122$-$169	& $-$2.09 (0.11)	& 17.4 (1.7)	& 7.82 (4.47)	& 4.08 (0.65) 	\\
173$-$223	& $-$2.14 (0.19) 	& 16.4 (2.7)	& 5.05 (3.75)	& 2.59 (0.72)	\\
 \hline
\end{tabular}
\end{center}
\begin{minipage}{0.4\textwidth}
\vspace{-6pt}
\hspace{-20ex}
\scriptsize{$^*$ Flux is calculated in 8$-$150~keV.}
\end{minipage}
\vspace{-10pt}
\label{tab:spec_result}
\end{table}

The average energy flux over the entire enhancement is $(6.5\pm2.4)
\times 10^{-8}$\,erg\,cm$^{-2}$\,s$^{-1}$ (in 8$-$150~keV), of which the
blackbody component accounts for 19\%.  
We estimate a total
isotropic emitted energy of $2.9 \times 10^{40}D_5^{~2}$\,erg for the
entire persistent emission (8$-$150~keV) during the enhancement.

To investigate the evolution of the blackbody component and of the
source's spectral properties in general, we divided the enhanced
emission period into three time intervals of $\sim$50~s each:
74$-$117~s, 122$-$169~s, and 173$-$223~s after the trigger time. 
We employed
the same set of photon models as the time-integrated analysis
described above.  The first spectrum was best fit by a single power
law with no evidence of a blackbody or any curvature.  The second and
third spectra, on the other hand, were best described by power law +
blackbody models.  In the second spectrum (the peak of the
enhancement) the additional blackbody component was statistically most
significant, and remained significant in
the third spectrum as well ($\Delta$C-stat = 42.9 and 15.3, corresponding to 6.2$\sigma$ and 3.5$\sigma$ improvements, respectively).  
The ratio of the blackbody flux to the
total flux (8$-$150~keV) was found to be 34\% in both intervals. The
indices of the underlying power-law component, and the blackbody
temperature also remained constant, at $\sim$$-2.1$ and $\sim$17~keV,
respectively (within uncertainties; see also Table 1), while the
power-law amplitude tracked the photon flux.

\subsection{Phase-Resolved Spectral Analysis}

We performed spin-phase-resolved spectral analysis of the pulsed
enhanced emission, as follows: we co-added the burst-free spectrum of
each pulse (in \trig+122 to 223~s, corresponding to second and third
time-resolved spectra) using TTE data and extracted a phase-maximum
spectrum (between phases 0.75$-$1.25 in Figure \ref{fig:pptte}) and a
phase-minimum spectrum (between phases 0.25$-$0.75). The spin phase
for each photon was calculated using barycentered times, as was done
for the timing analysis. We calculated the background spectrum, from
the burst-free interval at \trig~to \trig+60~s.

The spectra of both the phase minimum and maximum were adequately
fitted with power law + blackbody models, where we kept the power-law
indices and the blackbody temperatures linked.  The values of the
linked parameters found in the fit were consistent (within 1$\sigma$)
with those of the time-integrated spectra (see
Table~\ref{tab:spec_result}). 
The contributions of the blackbody flux to
the total flux were (52$\pm$18)\% and (35$\pm$18)\%, in the
phase-maximum and phase-minimum spectra, respectively.
This hints that the blackbody component was more significant in the phase-maximum spectrum than in
the phase-minimum spectrum.

\section{Summary and Discussion}\label{sec:sum}

We report here the discovery of coherent pulsations in the persistent
hard X-ray emission from \sgr~in the {\it Fermi}/GBM data lasting
$\sim$150~s. Coherent pulsations with a 55\% RMS pulse fraction have
never been detected in the persistent emission at these high energies
from a magnetar as yet. 
These pulsations were detected only at
the onset of a major bursting episode and were not directly related to
a major burst or flare from the source. 
Previously, intermediate flares with pulsating tails were observed
from SGR\,1900+14 \cite{ibr01,len03} and very recently from
\sgr~($\sim$6~hours after \trig)\cite{mer09}.  Thermal components
were also found in the decaying tails of intermediate events from
SGR\,1900+14 with much lower blackbody temperatures of $\sim$2~keV
\cite{len03}.  The thermal component of the enhanced emission we
report here is hotter (17~keV), exhibits a strong dependence of the
pulse profile with energy with a very high RMS pulsed fraction (up to
55\%), and is clearly not associated with a decaying event tail.
Energetically, however, the fluence of this enhanced emission is
comparable to that of tail emission of the intermediate flares from
SGR\,1900+14.

Our timing analysis showed that the detection of pulsations is most
significant in the 120$-$210~s interval after trigger. We find that
the spectrum requires a blackbody component along with a power law
between 122$-$223~s, which is consistent with the time interval of the
most significant detection of pulsations. Moreover, as determined by
the energy dependent pulse profiles and RMS pulsed fractions, we find
that the high-energy pulsations are most significant in the
50$-$74~keV range.  Strikingly, the blackbody component of the
enhanced persistent emission spectrum peaks at around 51~keV (i.e.,
the Wien peak of 17~keV, see Figure~\ref{fig:nufnu}). These two
independent pieces of evidence lend strong support for a blackbody
radiation component to account for the curvature in the spectrum of
the enhanced emission.

In our spin-phase-resolved spectral analysis, we find that the
blackbody flux to the total emission is (52$\pm$18)\% and
(35$\pm$18)\% in the phase-maximum and phase-minimum spectra,
respectively. This also suggests that a major contribution to the
observed pulsations is from the blackbody component. If we assume a
surface hot-spot during this pulsating interval, then the best-fit
blackbody corresponds to an effective radiating area (as projected on
the plane of the sky, far from the star) of $S_\infty =\pi D^2
F/(\sigma T^4) \approx 0.046D_5^{~2}\;{\rm km}^2$, where $T \approx
2\times 10^8\;^\circ$K ($kT \approx 17\;$keV) is the observed
(gravitationally redshifted) temperature. We have used here the
blackbody flux at the peak of the pulsations (i.e., phase maximum; $F
\approx 5\times 10^{-8}\;{\rm erg\;cm^{-2}\;s^{-1}}$) where the
hot-spot is expected to be relatively close to face-on, in order to
minimize the effects of projection and gravitational lensing by the
neutron star, so that $S_\infty$ would be relatively close to the
physical area, $S$, of the hot spot on the neutron star surface. For a
circular hot-spot, this corresponds to a radius of $\sim$120$D_5$\,m.

The internal field of magnetars can be significantly stronger and more tangled than the external dipole field.
As the internal field twists the stellar crust, the magnetosphere also becomes twisted, possibly in a complex manner~\cite{tho02}.
If we identify the inferred size of the hot-spot as the size of a twisted region on the stellar surface,
the rate of energy dissipation is expected to be $L_{\rm d} \ll 10^{38}\;{\rm erg\;s^{-1}}$ for global stability case \cite{bel07}.
This is inconsistent with the observed luminosity of the spot, $L_{\rm d} \sim 10^{38}D_5^{~2}\;{\rm erg\;s^{-1}}$ 

It may be possible for the magnetic twist to grow to a global
instability level during a highly active bursting period due to
frequent starquakes \cite{bel07}.
As the magnetosphere untwists, a large amount of energy must be
dissipated \cite{lyutikov}.  A small ``trapped fireball'' -- plasma
of $e^\pm$ pairs and photons confined by a closed magnetic field
region -- could then potentially account for the inferred hot-spot,
and in particular its roughly constant temperature and size. Confining
a ``fireball" of energy at least comparable to that emitted by the
observed blackbody component, $E_{\rm iso,BB} \approx 5.6 \times
10^{39}D_5^{~2}\;$erg, within a region of radius $a \sim 120D_5\;$m
requires $E_B(a) = \frac{1}{6}a^3B^2 > E_{\rm iso,BB}$ or $B \gtrsim
 1.4 \times
10^{14}D_5^{-1/2}\;$G. This is consistent with the surface dipole
field of $B\approx 2.2\times 10^{14}\;$G inferred from the measured
$P\dot{P}$ \cite{cam07}. Therefore, a sufficiently small closed
magnetic loop anchored by the crust could provide the required
confinement.
Moreover, the blackbody emission is expected to be accompanied by non-thermal,
high-energy radiation produced by collisionless dissipation, with 
the luminosities comparable to the blackbody components \cite{bel07}. This is in
good agreement with our observations of \sgr.

In conclusion, the area of the blackbody emitting region observed here is the smallest ``hot spot" measured for a magnetar, which likely arises from magnetically confined hot plasma on the neutron star surface, possibly caused by the gradual dissipative untwisting of the magnetosphere \cite{lyutikov}. If the total radiated energy was initially confined to the inferred extremely small size of the enhanced emission region (as in a mini ``trapped fireball'' scenario), this would indicate a very large magnetic energy density, similar to the ``trapped fireball'' model for the tails of SGR giant flares. 
The observed enhanced emission that we report here is much less energetic than a giant flare tail, while its energy is comparable to the tail energy of intermediate events and at the high end of typical SGR bursts.  
Despite some distinct properties, the enhanced emission of \sgr~carries various flavors of all three SGR phenomena, and thus it is most likely related to the very pronounced bursting activity that immediately followed it.

\bigskip 
\begin{acknowledgments}

 This publication is part of the GBM/Magnetar Key Project
(NASA grant NNH07ZDA001-GLAST, PI: C. Kouveliotou). We thank
G.L. Israel and A. Tiengo for providing the precise spin ephemeris and
source distance, respectively, prior to their publication. YK and EG
acknowledge EU FP6 Transfer of Knowledge Project ``Astrophysics of
Neutron Stars'' (MTKD-CT-2006-042722). EG acknowledges partial support from Turkish Academy of Sciences. JG gratefully acknowledges a
Royal Society Wolfson Research Merit Award. ER-R thanks the Packard
Foundation for support. 

\end{acknowledgments}

\bigskip 

\end{document}